\documentclass{ieee}
\usepackage{times}

\usepackage{boxedminipage}
\usepackage{epsfig}
\usepackage{psfrag}
\usepackage{subfigure}
\usepackage{graphicx}
\usepackage{comment}
\usepackage{afterpage}
\usepackage{cite}
\usepackage{amsthm}
\usepackage{mathenv}
\usepackage{mdwtab}
\usepackage{wrapfig}
\usepackage{graphics}
\usepackage{caption}
\usepackage{amsmath}
\usepackage{url}
\usepackage{multirow}

\pdfoutput=1

 \def\fixme#1{\typeout{FIXME in page \thepage : {#1}}\bgroup
\color{red}{FIXME - [{#1}]}\egroup}

\newcommand{\setfloattype}[1]{\def\@captype{#1}}

\newcommand{\reminder}[1]{ [[[ \marginpar{\mbox{\bf{   $\Leftarrow$}}} {\bf #1} ]]] }

\begin{document}

\title{
%
%
S3A: Secure System Simplex Architecture for \\ Enhanced Security 
of Cyber-Physical Systems
\thanks{This work is supported in part by a grant from Rockwell
Collins, by a grant from Lockheed Martin, by NSF CNS 06-49885 SGER, NSF CCR
03-25716 and by ONR N00014-05-0739. Opinions, findings, conclusions or
recommendations expressed here are those of the authors and do not necessarily
reflect the views of sponsors.}
}

\author{
\normalsize
Sibin Mohan, Stanley Bak, Emiliano Betti, Heechul Yun, Lui Sha and Marco
Caccamo\\
\normalsize
Dept. of Computer Science, University of Illinois at Urbana-Champaign,
Urbana IL 61801\\
\normalsize
[sibin,sbak2,ebetti,heechul,lrs,mcaccamo]@illinois.edu\\
\vspace*{3.0\baselineskip}
}

\pagestyle{plain}
\setcounter{page}{1}

\maketitle


\sloppy
\begin{abstract}
Until recently, cyber-physical systems, 
especially those with safety-critical properties that manage critical
infrastructure ({\em e.g.} power generation plants, water treatment facilities, {\em etc.}) were considered to be
invulnerable against software security breaches. The recently discovered
`W32.Stuxnet' worm has drastically changed this perception by demonstrating that
such systems are susceptible to external attacks.
Here we present an architecture that enhances the security
of safety-critical cyber-physical systems despite the presence
of such malware.  
Our architecture uses the property that control systems have deterministic
execution behavior, to detect an intrusion within {\em $0.6\ \mu s$} while
still guaranteeing the safety of the plant. We also show that
even if an attack is successful, the overall
state of the physical system will still remain {\em safe}. Even if the operating
system's administrative privileges have been compromised, our architecture will still be
able to protect the physical system from coming to harm. 

\end{abstract}





\section{Introduction}
\label{sec:intro}

Many systems that have safety-critical requirements such as power plants,
industry automation systems, automobiles, {\em etc.} can be classified as
{cyber-physical systems} (CPS) -- {\em i.e.} a tight combination of, and
co-ordination between, computational and physical components. Typically the `cyber' side
aids in the monitoring or control of the physical side. These systems (or parts of
them) have stringent safety requirements and require deterministic operational
guarantees since a failure to meet either could result in physical harm to the
system, the environment or even humans.

Such systems have traditionally been considered to be extremely secure since
they {\em (a)} are typically not connected to the Internet; {\em (b)} use
specialized protocols and proprietary interfaces (`security through obscurity')
{\em (c)} are physically inaccessible to the outside world and {\em (d)}
typically have their control code executing on custom hardware such as
specialized processors or programmable logic controllers (PLCs).

This misconception of ironclad security in such systems, however, has recently
been exposed when the `{\em W32.Stuxnet}' worm (henceforth referred to as 
just `Stuxnet') targeted and successfully infiltrated a Siemens WinCC/PCS7
control system \cite{stuxnet:cert10,stuxnet:cert10-1,stuxnet:cert10-2}. Not
only was it able to bypass all the security (digital as well as physical) techniques 
but it also reprogrammed the PLC that controlled the main
system. It modified the operating frequencies sent from the PLC thus {\em
causing physical damage to the system}\cite{stuxnet:motorexamplesymantec11}.

In this paper, we specifically address the problem of security for physical
control systems. Compared to general-purpose techniques, our work is
different in that we focus on domain-specific characteristics of these
cyber-physical systems and in particular, their {\em deterministic real-time
nature}. We introduce an overall system architecture where an isolated and
trusted hardware component is leveraged to enhance the security of the
complete system. We present a novel intrusion detection mechanism that monitors
context-specific side channels on the main CPU -- in this particular paper we
use the {\em deterministic execution time} of the control system as the main
side channel for this purpose \footnote{We elaborate on other potential
side-channels in Sections \ref{subsec:other_channels} and \ref{sec:concl}.} 

Program execution inherently includes variance due to a variety of features,
{\em viz.} complex control flow (branches, unbounded loops, {\em etc.}),
hardware features (caches, branch predictors, bus contention, {\em etc,}),
system effects (OS noise compiler optimizations, network traffic, interrupts,
{\em etc.}) and so on. Attackers often use these characteristics and other
vulnerabilities in the system to their advantage. Existing mechanisms
\cite{iyer2008,hardwaresecurity:deng2010} work well in detecting and preventing
problems, but either require custom configuration of reconfigurable hardware for each type of
checking mechanism \cite{hardwaresecurity:deng2010} or enforce run-time
monitoring and constraints on access to data by fine-grained checks on what
instructions/programs are allowed to access \cite{iyer2008}. Either way, there
is a distinct need to know more details about the program and data semantics. 
Typically good security involves one or more of the following principles: {\em
(i)} knowledge/use of control semantics; {\em (ii)} details about
program and data semantics and checking mechanisms; {\em(iii)} hardware-enabled
trust/protection; {\em (iv)} externally monitor-able information ({\em e.g.
}real-time execution time profile in our case) and {\em (v)} robustness/fault-tolerance mechanisms. 

Hence, we present the {\em Secure System Simplex Architecture} (S3A) to improve
the security of cyber-physical systems that uses (i), (iii), (iv) and (v) from
above as follows: a combination of knowledge of high-level control flow, a
{\em secure co-processor} implemented on an FPGA \footnote{Can be a trusted
processor/ASIC/unwritable FPGA in the final implementation}, deterministic
execution time profiles and System Simplex \cite{bak_simplex09,sha94}. S3A
detects intrusions that modify execution times by as low a value as $0.6 \mu s$ on our test control system. 
With S3A, we expand the definition of `correct system state' to include not just
the physical state of the plant but also the {\em cyber state}, {\em i.e.} the state
of the computer/PLC that executes the controller code. This type of security is
hard for an attacker to overcome by reverse engineering the code or the system
especially since it involves {\em absolutely no changes to the source
code/binary}. Even if an infection occurs and all of the security mechanisms
are side-stepped (such as gaining access to the administrative privileges or the
replication of our benevolent side channels), the trusted hardware component
(secure co-processor) and the robust Simplex mechanism will still prevent the
physical system from coming to harm, even from threats such as Stuxnet.

It is important to note that S3A is a {\em system-level solution that can
integrate multiple different solutions} to achieve security and
safety in this domain. While we picked some mechanisms (execution time, Simplex,
{\em etc.}), other good concepts and architectures
\cite{iyer2008,hardwaresecurity:deng2010} can also be integrated to
make the system that much more secure and robust. 

The {\bf main contributions} of this paper are as follows. We present the {\em
Secure System Simplex Architecture} (S3A) where,
\begin{enumerate}
\item a trusted hardware component provides oversight over an
untrusted real-time embedded control platform. This design 
provides a guarantee of plant safety in the event of successful infections. Even if an
attacker gains administrative/root privileges she cannot
inflict much harm since S3A ensures that the overall system (especially the
physical plant) will not be damaged.

\item we investigate and use of {\em context-dependent side
channels for intrusion detection}. These side channels, monitored by the
trusted hardware component, qualitatively increase the
difficulty faced by potential attackers. Typically side-channel communication is
used to break security techniques but we use them to our advantage in S3A. In
this paper, we focus on side-channels in the context of CPU-controlled real-time
embedded control systems.

\item we build and evaluate an {\em S3A prototype} for an inverted
pendulum plant and discuss implementation efforts and the construction of
side channel detection mechanism for {\em execution time-based side
channels} using and FPGA in the role of the trusted hardware component. The side
channel approach is shown to detect intrusions significantly faster than earlier
plant-state-only detection approaches.
\end{enumerate}
While intrusion detection is a broad area in computer security, our
approach takes advantage of the real-time properties specific to embedded
control systems. 
Also, most of the existing side-channel techniques/information (timing, memory,
{\em etc.}) have traditionally been used to break the security of systems.
This paper, proposes a method to turn it around so that these pieces of
information are now used for {\em increasing} the security of the system.
Also, such techniques have not been used before with the perspective of
safety-critical control systems -- hence we believe that this paper's
contributions are really novel.

We also believe that our approach is generalizable to PLC
and microcontroller-based CPS. Our justification is twofold; such systems {\em
(i)} have stringent requirements for correct operation, {\em i.e.}
the physical state of the plant must be kept safe under all conditions and {\em
(ii)} often require the controller process to run in a deterministic
manner.

\subsubsection*{Assumptions:} 
The important assumptions for the work presented in this paper are:
\begin{itemize}
\item the system consists of a set of periodic, {\em real-time} tasks with
stringent timing and deadline constraints managed by a real-time scheduler; such
systems typically do not exhibit complex control flow, do not use dynamically
allocated data structures, do not contain loops with unknown upper bounds, don't
use function pointers, {\em etc.} -- in fact, they are often designed/developed
with simplicity and determinism in mind 

\item the hardware component must be trusted and can only be accessed by
authorized personnel/engineers -- this is not unlike the RSA encryption
mechanism where the person holding the private key must be trusted
\item the systems we describe are rarely updated and definitely not
in a remote fashion (unlike,say, mobile embedded devices) -- see Section
\ref{sec:simplex} for details.
\end{itemize} 
{\bf Note:} Our techniques are not specific to attacks mentioned in Section
\ref{sec:stuxnet} and tackles the broader class of security breaches of controllers
in safety-critical CPS.

This paper is organized as follows: Section \ref{sec:stuxnet}
reviews breaches in safety-critical systems while Section
\ref{sec::attack_model} discusses the attack models that affect our work.
Section \ref{sec:simplex} provides a background of System-Level Simplex. 
Section \ref{sec:security}
presents the Secure System Simplex Architecture and implementation details.
Section \ref{sec:results} discusses the evaluation of
the system. Section \ref{sec:limitations} discusses 
 the limitations of our approach. Finally, related work is reviewed in Section
 \ref{sec:related} and Section \ref{sec:concl} concludes the paper and also
 presents some ideas for future work.
\section{Motivation}
\label{sec:stuxnet}

Many control systems attached to critical infrastructure 
systems have traditionally been assumed to be extremely secure.
The chief concern with such systems is related
to {\em safety}, {\em i.e.} to ensure that the plant's operations remain within a
predefined safety envelope. ``Security'' was attained by restricting external
access to such systems -- they were typically not connected to the Internet and
only a few people were granted access to the computers that controlled the
systems. Also, since parts (or even all) of the control code executed on
dedicated hardware (PLCs for instance), they were considered to be secure as well. 

\subsection{Stuxnet}
\label{subsec:stuxnet}

The {\em W32.Stuxnet} worm attack
\cite{stuxnet:cert10,stuxnet:cert10-1,stuxnet:cert10-2} overturned all of the
above assumptions. It showed that industrial control systems could now be
targeted by malicious code and that {\em not even hardware-based controllers were safe.}
Stuxnet employed a really sophisticated attack mechanism that took control of
the industrial automation system executing on a PLC. It took control of the
system and operated it according to the attacker's design. It was also able to
{\em hide these changes from the designers/engineers who operate the system}. To
achieve these results, Stuxnet utilized a large number of complex methods 
the most notable of which was the {\em first known PLC
rootkit}\cite{stuxnet:symantecdossier11}.
In fact, Stuxnet was present on the infected systems for a long time before it
was detected -- perhaps even a few months. Also, from all the information that
is available about the original attack, it seems that the worm made its way to
the original system through an infected USB stick. In this section we will focus
on the real target of Stuxnet -- the control code that manages the plants and
the implications of such an attack.

Stuxnet had the ability to {\em (a)} monitor blocks that were exchanged
between the PLC and computer, {\em (b)} infect the PLC by replacing legitimate
blocks with infected ones and {\em (c)} hide the infection from designers. 
This results in the operators being unaware of the infection, since the
information that they receive (supposedly from the PLC) shows everything to be
operating correctly. The PLCs are used to communicate with and control
`frequency converter drives' that manage the frequency of a variety of motors. The
malicious code in the infected PLC affects the operational frequency of these
motors so that they now operate outside their safety ranges. {\em E.g.}, in one
instance, the frequency of a motor was set to $1410$ Hz, then $2$ Hz and then to
$1604$ Hz and the entire sequence is repeated multiple times -- the normal
operating frequency for this motor is between $807$ Hz and $1210$ Hz
\cite{stuxnet:motorexamplesymantec11}. Hence, in this instance, Stuxnet's
actions can result in {\em real physical harm to the system}.

{\bf Note:} In this work, our focus is not on preventing the original intrusion
or providing mechanisms to safeguard the Windows machines that were originally
infected. We intend to detect the infection of the control code (on a PLC in
this example, but could be any computer that runs it) and mainly safeguard
the physical system from coming to harm.

\subsection{Automotive Attack Surfaces}
\label{subsec:automotive}

Researchers from the University of Washington demonstrated how a modern
automobile's safety can be compromised by malicious attackers
\cite{auto:koscher2010,auto:checkoway2011}. They show how an attacker is able to
circumvent the rudimentary security protections in modern automobiles and
infiltrate virtually any electronic control unit (ECU) in the vehicle and
compromise safety-critical systems such as disabling the brakes, stopping the
engine, selectively braking individual wheels on demand, {\em etc.} -- all of
this, while ignoring the driver's inputs/actions. They were able to achieve
this due to the vulnerabilities in the CAN bus protocols used in many modern
vehicles. The attackers also show how malicious code can be embedded within the
car's telematics unit that will completely erase itself after causing the crash.

This example is important, since the authors showed that the safety-critical
components of the vehicle can easily be targeted, thus putting at risk the
humans in the car. One important facet to note is that the critical components
that were attacked, such as engine control unit, braking unit, {\em etc.} all
have stringent real-time properties. Hence, our techniques will work well in
detecting the intrusions in these safety-critical subsystems.

\subsection{Maroochy Wastewater Attack and other Examples}
\label{subsec:maroochy}

In $2001$, an erstwhile employee of a small town in Australia started issuing
radio commands (using stolen equipment) to sewage treatment facilities that he
had helped install, using stolen equipment
\cite{scada:abrams2008,scada:smith2001}. This resulted in a lot of environmental
damage. The attack was hard to track since the requisite alarms were not being
reported to the central computer and this computer couldn't communicate with the
pumping stations during the attacks. Initially the incidents looked like
anomalous, unintentional events. It was took a lot of analysis of the system to
understand that there was a malicious entity operating to cause these problems.

There have been numerous other attacks that infiltrated critical systems 
{\em e.g.} NRG generation plants \cite{scada:nerc2009}, canal
systems \cite{scada:mcmillan07}, medical devices \cite{medicalsecurity:li2011},
{\em etc.}

\subsection{Discussion}
\label{subsec:motivation_discussion}

As these examples show, safety-critical systems can no longer be considered to
be safe from security breaches. While the development of cyber security techniques can
help alleviate such problems, the real concern is for the control systems and
physical plants that can be seriously damaged -- often resulting in the
crippling of critical infrastructure. Hence, we propose {\em non-traditional
intrusion detection and recovery mechanisms} to tackle such problems. We 
use to our advantage the fact that the control codes running in a real-time
system tend to be deterministic in behavior, simple to implement and exhibit
strict timing properties.

For the rest of this paper, we will show how such intrusions can be detected and
the harmful effects mitigated by use of our {\em Secure System Simplex
Architecture} (S3A). Hence, {\em our aim is to identify, as quickly as possible,
that an infection has taken place and then ensure that the system (and its
physical components) are always safe}. {\bf Note:} as stated in the
introduction, our work does not aim to prevent the original infections since
that is a large problem that requires the development and implementation of multiple levels of
cyber security techniques/research. We focus on the aftermath of
the infection of control codes.
\section{Threat Model}
\label{sec::attack_model}

We deliberately will not delve too deeply into specific threat models,
since we believe that our techniques will work well for a broad class of attacks that modify the
execution behavior of embedded code in safety-critical systems. Attacks similar
to the ones mentioned in Section \ref{sec:stuxnet} can be caught by the
mechanisms presented in this paper. Hence, code could be injected by any of the
mechanisms described in that section -- as long as the malicious entity tries to
execute {\em any} code, we will be able to detect it. Hence, our threat model
\cite{hardwaresecurity:iyer2007} is quite broad and can detect attacks such as: {\em (a)} {\em
physical attacks}, {\i.e.} code injected via infected/malicious hardware; {\em
(b)} {\em memory attacks} where attackers try to inject malicious code into the
system and/or take over existing code; {\em (c)} {\em insider attacks} where the
attackers try to gain control of the application/system by altering all or part
of the program at runtime.

We will, instead, focus on what happens {\em after} attackers perform any of the
above actions in order to {\em execute their code}. Hence, we intend to show how our
architecture is able to quickly detect this and keep the system(s) safe 
particularly the {\em physical systems}. Since we don't care much about {\em
what} executes and are more concerned with {\em how long} something executes, our
``malicious entity'' is a little more abstract as explained later in Sections
\ref{subsec:malicious} and \ref{subsec:malicious_results}.

\section{System Simplex Overview}
\label{sec:simplex}

\begin{figure*}[tb]
\hspace{0.2cm}
\begin{minipage}[b]{\columnwidth}
\begin{center}
\includegraphics[height=3.5cm,width=8cm]{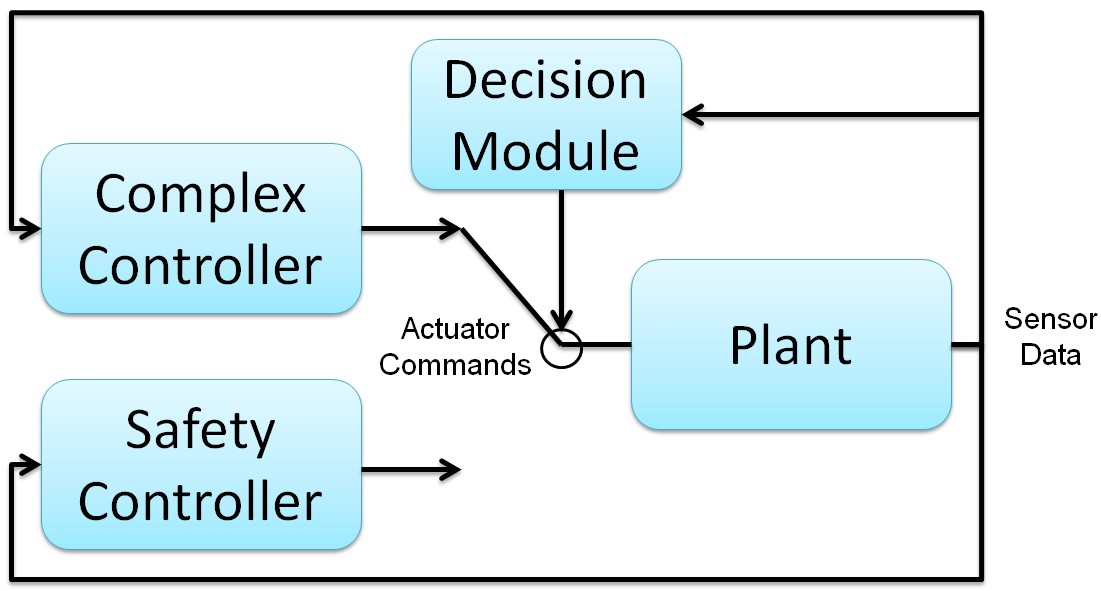}
\end{center}
\caption{Simplex Architecture}
\label{fig::simplex}
\end{minipage}
\hspace{0.6cm}
\begin{minipage}[b]{\columnwidth}
\begin{center}
\includegraphics[height=3.5cm,width=8cm]{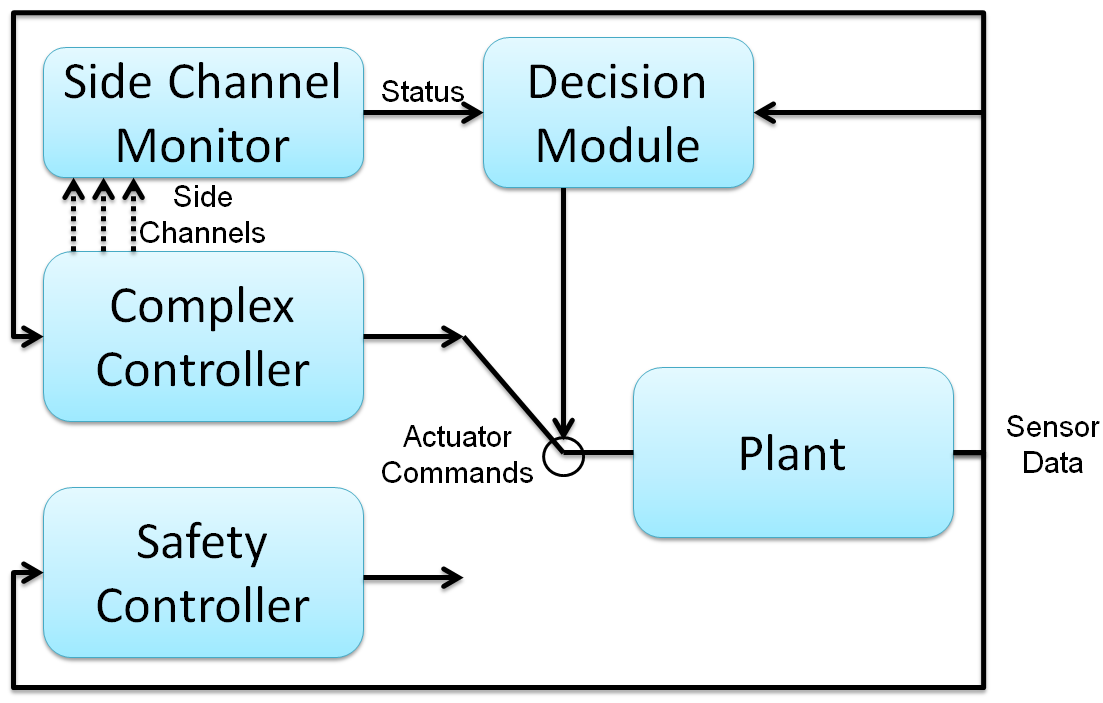}
\end{center}
\caption{S3A Architecture}
\label{fig::s3a}
\end{minipage}
\end{figure*}

The Simplex Architecture \cite{sha01} utilizes the idea of \textit{using
simplicity to control complexity} in order to safely use an untrusted subsystem
in a safety-critical control system. A Simplex system, shown in Figure
\ref{fig::simplex}, consists of three main components: 
\begin{enumerate}\itemsep1pt
\item[a.] under normal operating conditions the {\tt Complex Controller}
actuates the plant; this controller has high performance characteristics and is
typically unverifiable due to its complexity;
\item[b.] if, during this process, the system state becomes in danger of
violating a safety condition, the {\tt Safety Controller} takes over; 
\item[c.] the exact switching behavior is implemented
within a {\tt Decision Module}. 
\end{enumerate}
The advantage of the design is that
high-performance components can be used without the requirement that they be
fully verified. By maintaining a correct safety controller and decision module
the properties about the safety of the composite system can be guaranteed. Thus, 
even if the complex controller is upgraded, is faulty or becomes infected with
malware, we are still assured that the formal safety properties can never be violated and
the plant remains safe. The Simplex architecture has been used 
to improve the safety of a fleet of remote-controlled cars
\cite{crenshaw07_simplex}, pacemakers \cite{bak_simplex09} as well as 
advanced avionics systems \cite{simplex_autopilot}. 

Early Simplex designs had all three subsystems located in software -- at the
application-level. 
To guarantee complete system safety, however, other components such as any
middleware and the operating system need to behave correctly. This requirement was relaxed in 
\textit{System-Level} Simplex \cite{bak_simplex09} by performing
hardware/software partitioning on the system. In System-Level Simplex, 
the safety controller and the decision module are moved to a
dedicated processing unit (an FPGA) that is different from the the
microprocessor running the complex controller. We leverage this partitioning
technique in S3A.

\subsubsection*{Untrusted Controller:} 
One important question is the use of an unverified (and hence,
untrusted) complex controller in such systems. It is not that designers wish to
use unverified controllers in such systems. Most such controllers that
are intended to control anything but the simplest of systems are typically very
complex and hard to verify. This is especially true if they must also achieve
high levels of performance. Hence, there could be bugs and/or potential
vulnerabilities in the system that attackers could exploit. Even if we assume
that the controller is completely trusted, it can still be compromised (case in
point -- Stuxnet reprogrammed the controller in the PLC). Our
technique can protect against any such intrusion, be it in trusted or untrusted
controllers.

\subsubsection*{System Upgrades:}
Another issue is what happens if the system must be updated and
that process either {\em (a)} breaks the safety and timing properties of the
system or {\em (b)} introduces malicious code. This is particularly important if
such updates were to happen in a remote fashion. While these would be serious
issues in most general-purpose or even mobile embedded systems ({\em
e.g.} cell phones), it is not a problem for safety-critical systems. As
mentioned in Section \ref{sec:intro}, such systems are rarely updated, if at
all. Also, any updates have the following properties: 
\begin{enumerate}\itemsep1pt
\item[i.] updates are never performed remotely -- they are carried out by trusted
engineers;  
\item[ii.] most updates are minor in that they only tune certain parameters
and rarely, if at all, modify the control/timing structure of the code -- hence
they will not even modify the safety properties of the system and 
\item[iii.] any major changes will require extensive redesign, testing, {\em
etc.} -- hence the safety and real-time properties of the system  must then be
re-analyzed anyways. 
\end{enumerate}
One other important point is that the Simplex architecture can actually support
upgrades to the complex controllers \cite{simplex:seto98} in a safe 
manner.

Our application of Simplex in S3A, in this paper, has
several significant differences compared with earlier approaches. In the past,
the primary motivation to use Simplex was to aid in the verification of complex
systems. In this work, we instead apply Simplex to protect against malware that
has infected the complex controller. Another key difference is that previously
the decision module's behavior was determined completely by the physical state
of the plant. In this work, we widen the scope of the ``correct state'' by
using side channels from the computational part of the system, such as the
timing properties of executing real-time tasks, in order to determine when to perform 
the switching. The Simplex decision module is now monitoring \textit{both}, the
physical system as well as the cyber state of the computational system.

\section{Integrated Framework for Security: Secure System Simplex Architecture (S3A)}
\label{sec:security}

We now present the {\em Secure System Simplex Architecture} (S3A) that prevents
damage from malicious intrusions in safety-critical systems as well as aids in rapid
detection through side-channel monitoring. In this section, we first
elaborate on the high-level logical framework of the architecture.
We then discuss aspects of the execution time-based side channels that we have
implemented in our S3A prototype and then follow it up with details on
how to implement such a system -- from the hardware aspects to the OS
modifications; from the timing measurements to the control system that we use
to show the effectiveness of our approach.

\subsection{High Level Architecture}
\label{ssec:logical_arch}

Figure \ref{fig::s3a} provides a high level overview of the system
architecture. There is a {\tt Complex Controller} that computes the control logic
under normal operations. The computed actuation command is sent to the plant
and sensor readings are produced and given to the controller to enable feedback
control. There is also a {\tt Decision Module} and {\tt Safety Controller} in
this architecture that are used not only to prevent damage to the plant in case of
controller code bugs (as with the traditional Simplex applications) but also to
prevent plant damage in the case of malicious actuation from attackers. We also
have a {\tt Side Channel Monitor} that examines the execution of the {\tt
Complex Controller} for changes in `expected' behavior (in this paper it
monitors the execution time of the {\tt Complex Controller} to see if there is
any deviation from what is expected). If the information obtained via the side
channels differs from the expected model(s) of the system, the {\tt Decision
Module} is informed and control is switched to the {\tt Safety Controller} (and
an alarm can be raised). The types of side channels we can consider in a
CPU-based embedded system include the execution time profiles of tasks, the
number of instructions executed, the memory footprint and usage pattern or even
the external communication pattern of the task. We will discuss timing side
channels in more detail in the Section \ref{subsec:timing_channels} and
elaborate on the viability of the others in Sections \ref{subsec:other_channels}
and \ref{sec:concl}.

This approach is qualitatively more difficult to attack than a  typical control
system. An attacker not only has to compromise the main system, but she also has
to replicate all the side channels that are currently being monitored. If the
timing of the task execution is being monitored then the attacker must replicate
the timing profile of a correctly-functioning system. If the cycle count is being
observed, her attack must also be sure to execute for a believable number of
instructions. Even if all the side channels match the expected models, the
Decision Module will still monitor the plant state and, when malicious actuation
occurs, prevent system damage.

The effectiveness of the side channel early-detection methodology depends on two
factors. First, the constructed model of each side channel should restrict valid
system behavior (not easily replicable). Second, the side channel itself must be
secure (not easily forgeable). These factors are implementation specific and
will be discussed later in Section \ref{subsec:framework}.

\subsection{Timing Side Channels}
\label{subsec:timing_channels}

In this paper, we intend to secure a real-time embedded system. Therefore,
we assume that the system has typical real-time characteristics, {\em i.e.} the
system is divided into a set of periodic tasks managed by a real-time scheduler.
Each task has a {\em known execution time} and each task periodically activates
a job.

The monitoring module maintains a real-time timing model of the system.
Violations of this timing model occur when the,
\begin{itemize}\itemsep1pt
  \item[i.] job execution time is too large;
  \item[ii.] job execution time is too small; 
  \item[iii.] job activation period is too large; or
  \item[iv.] job activation period is too small.
\end{itemize}
Additionally, the monitoring module needs to examine the execution of the
{\em idle task}. This prevents a malicious attacker from allowing the real-time
task to execute normally and perform malicious activity during idle time.
Finally, the monitoring module should be cognizant of the system activities that
may result in timing perturbations.

In our prototype, we address two of these timing side channel
requirements: {\em monitoring the control task and the idle task}. For
rapid prototype development, we eliminate system noise (disable interrupts) while our
control task is running to obtain a predictable timing
environment\footnote{Details in Section \ref{subsec:os}.} rather than patching
system interrupts in order to receive their timing information. In a real-time
system the interrupts would be predictable and scheduled deterministically --
hence we would be able to monitor them as well as other tasks. This addition,
however, could be made to our prototype in the future. 

Execution times of the various real-time tasks in such systems are anyways
obtained as part of system design by a variety of methods \cite{wilhelm07}. 
There is no extra effort that we have to perform to obtain this information. The
worst-case, best-case and average-case behavior for most real-time systems is
calculated ahead of time to ensure that all resource and schedulability
requirements will be met during system operation. We use this knowledge of
execution profiles to our advantage in S3A.

\begin{table}[b]
\begin{center} 
\begin{tabular}{|l|l|}
\hline
Component & Details \\
\hline
\hline
  Inverted Pendulum & Quanser IP01 \\
\hline
  FPGA & Xilinx ML505 \\
\hline
  Computer with Controller & Intel Quad core 2.6 GHz \\
\hline
  Operating System & Linux kernel ver. $2.6.36$ \\
\hline
  Timing Profile & Intel Timestamp \\
  					  & Counter (rdtsc)\\
\hline
\end{tabular}
\end{center}
\caption{S3A Prototype Implementation Details}
\label{table:s3a_implementation}
\end{table}

\subsection{Other Potential Time-based Side Channels}
\label{subsec:other_channels}

In the assumed context of predictable real-time embedded control systems,
several other side channels are available as part of the cyber state. These
include the {\em task activation periodicity, memory footprint, bus access times
and durations, scheduler events, etc.}. Each of these is a candidate for {\em
benevolent side-channels} that can be monitored to detect infections and would
have to be individually detected and replicated by an attacker to maintain 
control in an infected system, thus qualitatively increasing the difficulty for
future attackers.

Additionally, the specific side channels used may vary depending on the type
of system. {\em E.g.}, in this paper, we focus on CPU-based real-time control
systems. Other systems, {\em e.g.} PLC-based systems, would likely need to
either monitor the side channels using different mechanisms or utilize a
completely different (or additional) sets of side channels.

\subsection{Implementation}
\label{subsec:framework}

We now describe a prototype implementation of S3A that we have created. 
%
The technical details of the prototype are listed in Table
\ref{table:s3a_implementation}. We will elaborate on key aspects of our
implementation in detail in the upcoming subsections. First, a hardware
component overview is provided in Section \ref{ssec:impl_overview}. Then, the
inverted pendulum hardware (our example `safety-critical control system') setup
is described in Section \ref{subsec:ip}. The methodology for timing measurements
of the control code is described in Section \ref{subsec:timing} and the
methodology for timing-variability (`malicious code') tests is presented in
Section \ref{subsec:malicious}. Section \ref{subsec:os} gives essential details about
the operating system setup during the measurements. Finally, Section
\ref{subsec:detection} describes the specific design of the {\tt Decision
Module} and the timing {\tt Side Channel Monitor}.

\begin{figure}[b!]
\small
\centering
\includegraphics[width=0.9\columnwidth]{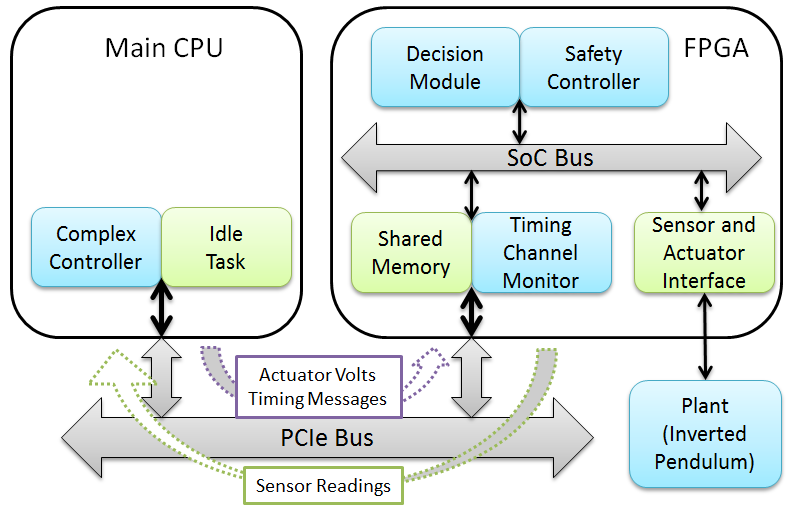}
\caption{S3A Implementation Overview}
\label{fig::s3aImpl}
\end{figure}

\subsubsection{Hardware Components}
\label{ssec:impl_overview}

A high-level hardware design of our prototype is shown in Figure
\ref{fig::s3aImpl}. The prototype hardware instantiates the logical Secure System Simplex
architecture previously described in Section \ref{sec:security} and shown in
Figure \ref{fig::s3a}. In our implementation, we run the {\tt Complex
Controller} on the main CPU. The Complex Controller communicates 
with a trusted hardware component, an FPGA in this case, 
to perform control of an inverted pendulum. Sensor readings are obtained by
the FPGA over the PCIe bus using memory mapped I/O. The
actuation command, in turn, is written to the memory-mapped region on the FPGA.
Additionally, timing messages in the form of memory-mapped writes are
periodically sent to the FPGA based on the state of execution (at the start/end
of the control task and periodically during the {\tt Idle Task}). This
creates a timing side channel that can be observed by a {\tt Timing Channel
Monitor} running on the FPGA. On the FPGA side, the {\tt Timing Channel Monitor}
will measure the time elapsed between timing messages from the {\tt Complex
Controller} to ensure that the execution conforms to an expected timing model.
The {\tt Decision Module} will periodically examine the output of the Timing
Channel Monitor, the actuation command from sent by the Complex Controller from
{\tt Shared Memory} on the FPGA, the actuation command from the locally-running
{\tt Safety Controller} and the state of the plant from a {\tt Sensor and
Actuator Interface} and decide which controller's actuation command should be
used -- the complex one on the CPU or the safe one on the FPGA. The actuation
command is then output back to the Sensor and Actuator Interface. The interface
then, through a digital-to-analog converter, actuates the {\tt Plant} -- in our
case, an inverted pendulum. The Sensor and Actuator Interface also periodically
acquires sensor readings through analog-to-digital converters and write their
values to both shared memory accessible by the Complex Controller and to memory
accessible by the trusted Decision Module and Safety Controller.

\subsubsection{Inverted Pendulum}
\label{subsec:ip}

\begin{figure}[h]
\small
\centering
\includegraphics[width=0.9\columnwidth]{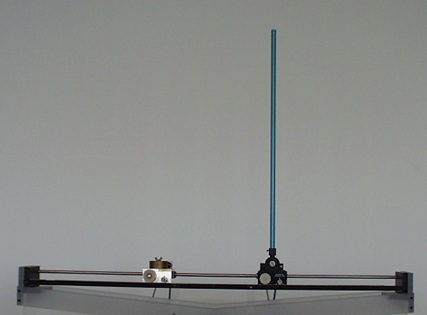}
\caption{An inverted pendulum control system maintains an upright rod along a one-dimensional track.}
\label{fig::ip}
\end{figure}

We used an inverted pendulum (IP) as the plant that was being controlled. An
inverted pendulum, like the one shown in Figure \ref{fig::ip}, is a classic
real-time control challenge where a rod must be maintained in an upright
position by moving a cart attached to the bottom of the pendulum along a
one-dimensional track. There are two sensors to measure both 
the current pendulum angle as well as the cart position on the track and there
is one actuator (the motor near the base of the pendulum) used to move the cart.
Two safety invariants must be met: {\em (1)} the pendulum must remain upright
(can not fall over) and {\em (2)} the cart must remain near the center of the
track. The specific inverted pendulum we used in our testbed was based on the
Quanser IP01 linear control challenge \cite{quanser_ip}.

Our setup varies slightly from an off-the-shelf version of the Quanser IP01.
The most important difference is that we needed to directly connect the sensors
and actuators to the FPGA hardware while the prebuilt setup requires a computer
to do the data acquisition. We modified the system to redirect the sensor values
and motor commands through an Arduino Uno microcontroller that communicates
directly with the S3A FPGA  through a serial cable. Although this change can
potentially introduce latency into the system, we did not observe any issues
with safely actuating the pendulum due to this small delay. The control code
that manages the IP executes on a computer (Section \ref{subsec:os} and Table
\ref{table:s3a_implementation}). This control code executes at a frequency of
$50$ Hz.

{\bf Note:}  The inverted pendulum has been used 
quite extensively in literature to be an appropriate example
of a real-time control system \cite{sha01,bak_simplex09}. Hence we believe it would
suffice to show an early prototype of our solutions. We are currently working
on applying these techniques to other real control systems in conjunction with
power vendors.

\subsubsection{Timing}
\label{subsec:timing}

The implementation of the complex controller for the inverted pendulum is fairly
simple with very few branches and most loops being statically decidable\footnote{This is
typical of most control code in safety-critical and real-time control systems
-- hence our implementation of the controller for the inverted
pendulum is also similar.}. Hence it is fairly easy to
calculate the execution time and number of instructions taken for such code. In
our framework, we utilized simple dynamic timing analysis \cite{wilhelm07}
methods to obtain an {\em execution profile} of the code. We used the Intel
{\em time stamp counter} ({\tt rdtsc}) \cite{rdtsc98} to obtain high resolution
execution time measurements for the control code.


The profile consisted of the `worst-case,' `best-case' and `steady-case'
numbers for the control code that was obtained by executing it multiple 
times on the actual computer where it would execute and measuring each
set of executions. `Steady-case' refers to the values obtained when the
execution time has stabilized over multiple, repeated executions -- {\em i.e.}
when the initial cold cache related timing dilation at the start of the
experiments no longer occur. 

The control code was placed in a separate function and
called in a loop. As part of our experiments, the loop was executed $1, 10, 100,
1,000, 10,000, 100,000$ and $1,000,000$ times. During each of these scenarios,
the total time of the loop as well as the times taken up during each individual
iteration was measured. From these traces we were able to determine the maximum
(worst-case), minimum (best-case) and steady-case values for the
execution time of the controller code. To reduce the noise from
instrumentation and overheads of the loops, function calls, {\em etc.} we used
the `dual-loop timing' method: {\em i.e.}, empty loops with only the
measurement instrumentation were timed as a 'control' experiment.
The execution times obtained for these instrumentation-only loops were subtracted
from the execution times for the loops with the control code. 

Interrupts (all interrupts including inter-processor interrupts) were disabled
during the timing measurements. To reduce the timing effects of the operating
system and other system issues we isolated our controller as best as we could
as we will describe in Section \ref{subsec:os}.

While we used simple measurement-based schemes for obtaining the execution
profile for the control code in this paper, it does not preclude the use of
other more sophisticated analysis techniques \cite{wilhelm07,mohan08} to obtain
better (and safer) timing estimates. This is especially true if the code is
more complex than the one used for the inverted pendulum. In fact, the better the
estimation methods, the better S3A will be able to detect anomalies and
intrusions.

\subsubsection{Execution Time Variation}
\label{subsec:malicious}

To mimic the effect of code modification on timing, we insert extra code into
the execution of the control loop function described in Section
\ref{subsec:timing}. Specifically, the extra code is a loop with a varying upper
bound ({\em i.e.} $1, 10, 100$) that performs multiple arithmetic operations
(floating point and integer). The idea being that the extra time/instructions
that execute will make it look like an intrusion has taken place. Our S3A system
will then detect the additional execution, raise an alarm and transfer
control to the simple controller on the FPGA.

{\bf Note}: As mentioned before, we are less interested in what kind of code executes
``maliciously'' because our detection scheme does not depend on this detail. We
only need to check whether whatever is executing has modified the timing profile of the system.

\subsubsection{System and OS Setup}
\label{subsec:os}

As stated in Table \ref{table:s3a_implementation}, we used
an off-the-shelf multi-core platform running Linux kernel $2.6.36$ for our
experiments. Since we use a COTS system there are many potential sources of
timing noise such as cache interference, interrupts, kernel threads and other 
processes that must be removed for our measurements to be meaningful.
In this section we describe the configuration we used to best emulate
a typical uni-processor embedded real-time platform.

The CPU we used is an Intel Q6700 chip that has four cores and each pair of
cores shares a common level two (last level) cache.
We divided the four cores into two partitions:
\begin{enumerate}\itemsep1pt
\item the {\em system partition} running on the first pair of cores (sharing one
of the two L2 caches) handles all interrupts for non-critical devices (e.g., the
keyboard) and runs all the operating system activities and non real-time
processes ({\em e.g.}, the shell we use to run the experiments); 
\item the \emph{real-time partition} runs on the second pair of cores (sharing
the second L2 cache). One core in the real-time partition runs our real-time
tasks together with the driver for the trusted FPGA component; the other core is
turned off so that we avoid any L2 cache interference among these two cores. 
\end{enumerate}

\subsubsection{Detection}
\label{subsec:detection}

\begin{figure}[b!]
\small
\centering
\includegraphics[width=0.8\columnwidth]{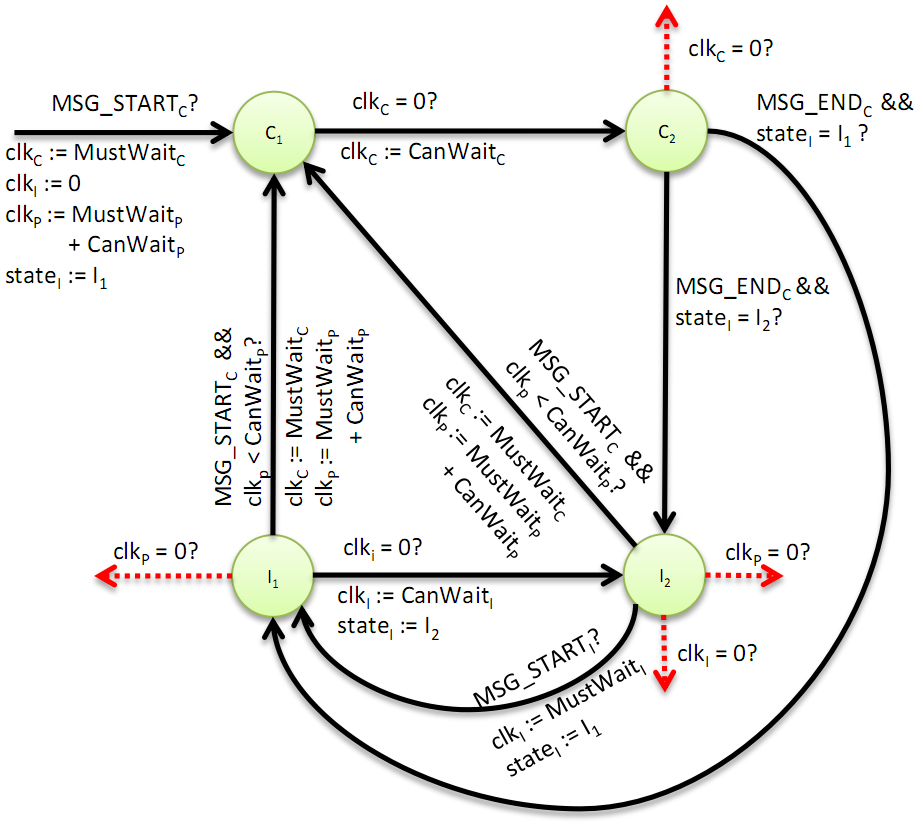}
\caption{FSM for Detecting Timing Model Violations}
\label{fig::detection}
\vspace{-2.0\baselineskip}
\end{figure}

In our implementation, detection of malicious code can occur in one of two ways.
The decision module observes both {\em (i)} the physical state of the plant 
(by traditional Simplex) as well as {\em (ii)} the computation state of the
system (based on timing messages; S3A). A violation of the physical model or the
computational model can trigger the decision module to switch control to the safety controller on the
FPGA. The physical model is monitored as described in previous work
\cite{sha01, bak_rtas2010, bak_iccps2011}. Based on a function of the track
position and pendulum angle, the decision module may choose to switch over to
the safety controller.

The computational system is also monitored for violations of the expected timing
model of the system. Both the control task and the idle task are monitored in
order to periodically send timing messages to the FPGA. The FPGA contains an
{\em expected timing model of the system that is a finite state machine} (FSM)
{\em running in hardware}. When timing messages arrive, or timers expire, the
finite state machine state can advance. If malicious code were to execute, it
would have a limited window of time to replicate the timing side channel before
it was detected by the Simplex module.

Generally speaking, monitoring the timing of a real-time system can be performed
by maintaining state about each task in the system. Each task would have a two
timers:
\begin{itemize}\itemsep0.5pt
\item[I.] the first would enforce the execution time of the task
\item[II.] the other will monitor periodic activation of the task 
\end{itemize}
Furthermore, a stack would be used to track task preemptions. Since typical
real-time systems use priority-based execution, all task switches will be
directly observable by the FPGA through task start/task end messages.

For our specific prototype, we implemented, in hardware, the finite
state machine on an FPGA as shown in Figure \ref{fig::detection}. In our system
there are two tasks: {\em (i)} the idle task and {\em (ii)} the controller task.
Since only a single task may be preempted (the idle task), we maintain a single variable as the call stack,
state$_I$. Three timers are present: clk$_C$ and clk$_P$ maintain the execution
time and period of the control task, clk$_I$ maintains the execution of the idle
task. In the figure, clk$_C$ ticks while the control task is running (states
C$_1$ and C$_2$) and clk$_I$ ticks while the idle task is executing (states
I$_1$ and I$_2$).  Clk$_P$ always ticks. The finite state machine is
parameterized with six values: MustWait$_C$, CanWait$_C$, MustWait$_I$,
CanWait$_I$, MustWait$_P$, and CanWait$_P$. These values are determined by the
minimum and maximum time permitted between timing messages. The MustWait time
indicates the minimum time that must elapse, whereas the CanTime indicates the
jitter permitted between different iterations of the loop. To say it another
way, MustWait is the minimum execution time of the task/idle loop/period,
whereas (MustWait + CanTime) is the maximum execution time.

In the finite state machine, initially the control task is running.
State C$_1$ is entered until clk$_C$ ticks from MustWait$_C$ to 0. Then state
C$_2$ is entered. If clk$_C$ ticks from CanWait$_C$ to 0 without the end task
message, the control task has executed for too long and a timing violation
occurs (indicated by the dotted arrow in state C$_2$). Once the end control task
message is received, the idle task begins to execute. Under normal operation,
the state will change between I$_1$ and I$_2$ several times, until the control
task is reactivated and state C$_1$ is again entered. Any messages that arrive
without explicit transitions in the timing FSM are interpreted
as errors in the prototype and trigger decision module to switch to the safety
controller. Additionally, the dotted transitions in the FSM are timing
violations that also trigger the decision module to take corrective action.

The FSM can also be used to tightly track the execution behavior of the code for
more sophisticated controllers, {\em e.g.} if the control code has many
branches, function calls, {\em etc.} For instance, when the control code reaches a branch
that affects the overall execution time, a message can be sent to the FSM about
which side of the branch was taken. The FSM can now use this information to
accurately track the execution of the program for all control constructs in the
code. 

\section{Evaluation}\label{sec:results}

In this section we evaluate the Secure System Simplex architecture -- first the
we present timing results that we obtained by analysis of the controller code
(Sections \ref{subsec:timing_results} and \ref{subsec:malicious_results}) -- these values are
used to form the profile of the execution behavior that is then used in
intrusion detection mechanism on the FPGA. We then present the details of the
intrusion detection in Section \ref{subsec:intrusion_results}.

\subsection{Timing Results and Execution Profile}\label{subsec:timing_results}

\begin{figure}[thb]
\small
\centering
\includegraphics[width=0.9\columnwidth]{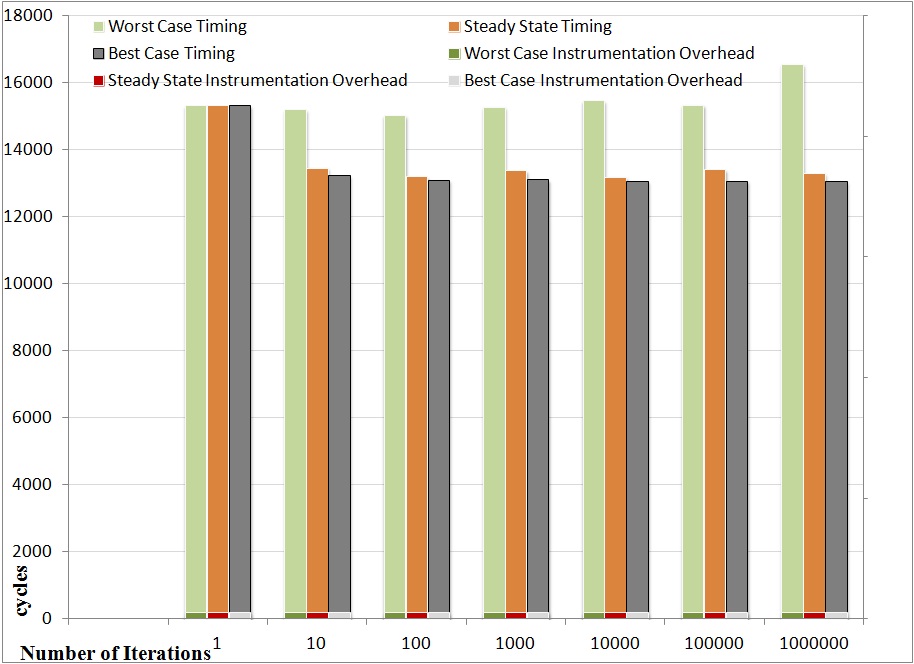}
\caption{Summary of the Timing Results}
\label{fig::results_summary}
\end{figure}

Figure \ref{fig::results_summary} shows a high level summary of the timing
experiments used to obtain the execution profile of the complex controller code
that executes on the computer (Figure \ref{fig::s3a}). We used dynamic 
timing analysis techniques to obtain the worst, best and steady state execution
times for this code. The x-axis represents the number of times the controller code was
repeatedly executed: from $1$ to $1,000,000$ in steps of $10$. The y-axis
represents the execution time in {\em cycles}. Each grouping of vertical lines 
represents the `worst-case', `steady-state' and `best-case' execution times for
that experiment. `Steady-state' refers to the execution time when successive
executions of the controller code resulted in the same execution time -- {\em
i.e.} the situation when the execution reached a steady state. This is compared
to the first few iterations, when cache and other hardware effects would result
in a higher variance in the execution time of the code -- the `worst-case'
numbers in the graph are usually from these first few iterations before the
system effects (in particular the cache) have settled down. This is the reason
why there exists a slightly larger difference between the worst-case and
best-case numbers.

Each vertical bar is split into two parts -- the lower part
shows the instrumentation overhead for that experiment\footnote{As
explained in Section \ref{subsec:timing}, we used dual-loop timing  techniques
to obtain the overheads due to the instrumentation.}, while the top is the
part that represents the pure timing for the control code only. 
We also see that the instrumentation overhead is almost the same across all
experiments -- oscillating between $260$ and $270$ cycles for all experiments.

As seen in the graph, the steady state and best-case values are very close, not
just within the same experiment, but across experiments. The largest difference
between the two is $360$ cycles for the $n=100,000$ experiment. This just shows that our assumption
that controller codes in safety-critical systems are simple and have little 
variability is valid. 
This lack of variability is also evident from the fact that the worst-case
execution cycles, across experiments, do not show much variance. The worst-case
values for the last experiment ($1,000,000$) has a slightly higher value of
$16,560$ and this can be chalked down to the initial cold cache and other system
effects. 

\begin{figure}[b]
\small
\centering
\includegraphics[width=0.9\columnwidth]{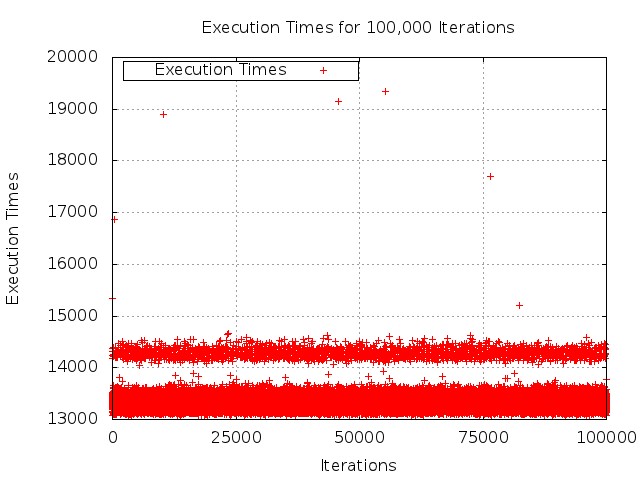}
\vspace{-0.5\baselineskip}
\caption{Execution Profile for 100,000 iterations}%
\label{fig::results_100000}
\end{figure}

Figure \ref{fig::results_100000} shows the execution profile for one timing
experiment in particular -- that of $100,000$ iterations. The
x-axis is the iteration number while the y-axis is the number of cycles for each iteration. As this
figure shows, the first few iterations take a little longer (around $17 K$
cycles) and then most of the execution stabilizes to within a narrow band of:
\begin{eqnarray*}
1,590\ cycles & = & 14,660 - 13,070 \\ 
i.e.\ \sim 0.6\ \mu s\ & at & \ 2.67\ GHz
\end{eqnarray*}
Hence, this band will define the `{\em accepted range}' of values that the FPGA
will use to check for intrusions. Any execution that changes the steady state
execution time by more than this narrow range will be caught by the FPGA. In
fact, the FPGA will catch variance in either direction -- {\em i.e.} an increase
as well as a decrease in execution time. 

The graph also shows that while the majority of execution times fall within a
small band at the lower end of the above mentioned range, some values
also fall into a narrow band at the top of the range ({\em i.e.} around the
$14 K$ value). This narrow band of increased execution times for some
experiments is due to latent system effects that we were not able to remove. The
main culprit is the last level cache that, in this architecture, uses a random
replacement policy. Hence, every once in a while a few of our controller's cache
lines are evicted by periodic kernel threads that we could not easily
disable (since we are running a COTS operating system) and these
iterations take a few hundred cycles extra (anywhere from $500-900$) to
execute. With a more predictable cache
replacement policy, like the ones used in hard real-time systems, we would not
see this behavior. To prove this theory we ran the same measurements on a
PowerPC that uses psuedo-LRU (Last Recently Used) cache replacement policy in
its last level and all the points are clustered into a single band. In fact,
with LRU, tasks would not evict each other cache's lines, unless the cache is
not big enough to fit them at the same time\footnote{If this is the case, we just
have to account for it, when we compute the execution time for each task.}

\begin{figure}[b]
\small
\centering
\includegraphics[width=0.9\columnwidth]{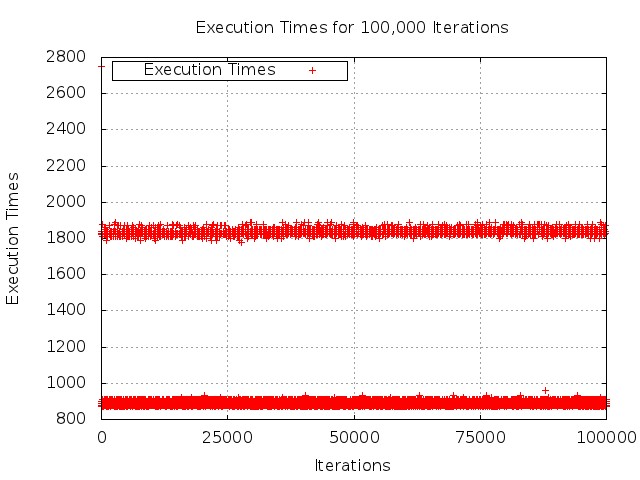}
\vspace{-0.5\baselineskip}
\caption{Execution Profile (100,0000) without FPGA}
\label{fig::results_100000_nofpga}
\end{figure}

Figure \ref{fig::results_100000} also shows a few sporadic experiments
exhibiting much higher execution times. Again, this is due to system effects, and in
particular, contention on the bus when communicating with the FPGA. As explained
in Section \ref{ssec:impl_overview}, the complex controller reads and writes
messages to and from the FPGA to control the pendulum and to send the timing
messages.
Many a time, while the complex controller is waiting for data from
the inverted pendulum (angle and track position) that arrives on the common
bus, the incoming messages
experience unpredictable delays. These delays are due to bus contention among
the FPGA and other peripherals sharing the same bus.

To prove that the communication with the FPGA was the cause
of these effects, we conducted timing experiments where the FPGA was switched
off and all calls to communicate with it (read/write) resulted in null function
calls. Figure \ref{fig::results_100000_nofpga} shows the results of these
experiments for the $100,000$ iterations point. This experiment highlights two
important points: 
\begin{itemize}\itemsep1pt
\item the random spikes at higher values no longer exist, thus showing that the
bus contention due to communication with the FPGA was the main cause of the
spikes
\item the same `double-band' of execution results also appears here; the
interesting fact is that the gap between the bands is almost exactly identical
to that of Figure \ref{fig::results_100000}, thus providing more evidence to the
fact that the cache (and its replacement policy) is the culprit.
\end{itemize}

Such issues could be avoided in an actual hard real-time system instead of the
COTS-based experimental setup that we use here. In fact, a hard real-time system
would use a more predictable bus, or other techniques \cite{rtBridge}, that
allows designers to bound I/O contention and avoid random spikes.

\subsection{Malicious code Execution Results}\label{subsec:malicious_results}

\begin{figure}[hbt]
\small
\centering
\includegraphics[width=0.9\columnwidth]{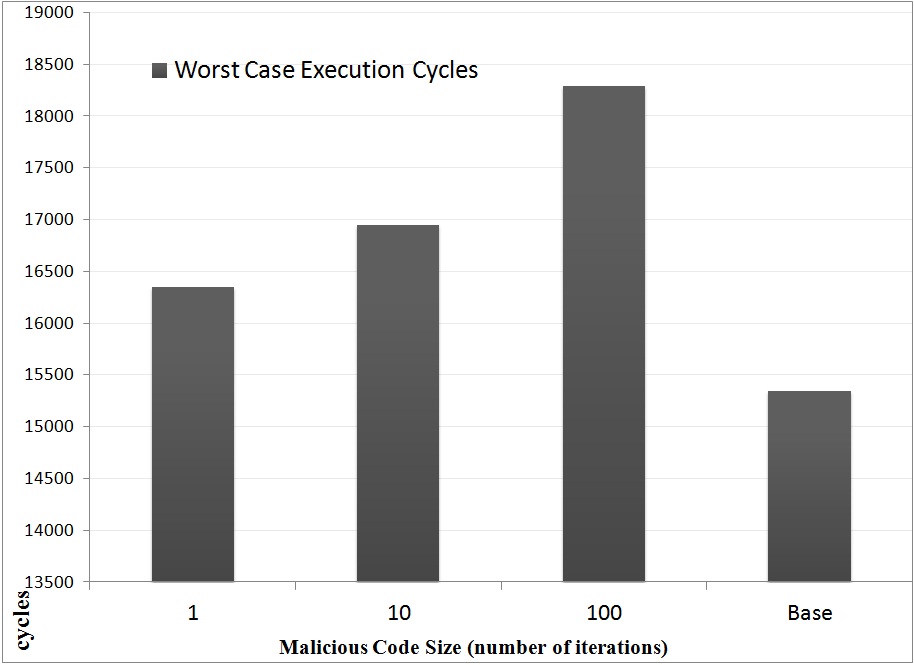}
\caption{Execution Cycles for Malicious Code execution}
\label{fig::results_malicious}
\end{figure}

We introduce ``malicious code'' by inserting extra instructions (Section
\ref{subsec:malicious}) -- {\em i.e.} a loop of variable size within the complex 
controller code. The upper bounds for the malicious loop are one of $1, 10, 100$
-- we stopped at the upper limit of $100$ since anything over this value would
put the execution of the ``infected'' control code over the real-time period of
the task. Also, as we will see soon, even these small additional increases in
execution times will be caught by the monitoring framework of S3A.

Figure \ref{fig::results_malicious} shows
the execution time (in cycles on the y-axis) taken up by the code for value of
the malicious loop values (x-axis). The final bar in the graph represents the
``base,'' {\em i.e.} the number of execution cycles taken up by the controller
code without any malicious loop. As expected, the values for the malicious code
increases significantly with each increase in the loop bound. 
Even the smallest sign of the presence of the malicious loop puts it outside of
the narrow range ($0.6 \mu s$) explained in Section \ref{subsec:timing_results}.
Hence, even this will be caught by S3A and control will be transferred over to
the simple controller executing on the FPGA.
{\bf Note:} Since we don't really care {\em what} executes as part of malicious
code and {\em intend to only catch variations in execution time}, we only
mimic the increased execution time effects by the methods discussed in this
section.

\subsection{Intrusion Detection}
\label{subsec:intrusion_results}


In this section, we describe the evaluation of our timing side channel intrusion
detection technique. First, we describe measurements of the timing of key
aspects of our overall architecture. Then, we demonstrate the {\em early
detection of malicious code execution} using the timing side channel approach
compared with monitoring the plant state only (say, by used of traditional
Simplex). The results of our intrusion detection measurements are summarized in
Table \ref{tab:intr_timing}.

Our first timing measurement was to check the {\em overhead of sending
timing messages} to the FPGA. Although the message itself takes time to
propagate through the PCIe bus to the FPGA, the CPU is not stalled during this time. By
using the time stamp counter we measured the overhead on the CPU for sending a
single timing message to be $130$ cycles ($50$ nanoseconds). This time is
extremely small and therefore each process could realistically send multiple messages
during a single iteration of each control loop to reduce the time an attacker
has to replicate the timing side channel. Another advantage of having multiple
timing messages per iteration is that if the program contains branches, we could
communicate to the FPGA timing monitor (at run-time) information about which
branch was taken thus allowing for tighter monitoring of the timing requirements
in the timing model FSM.

The second timing measurement we did was to quantify the {\em jitter of the
timing messages} through the interconnect going to the FPGA. We performed this
measurement by recording the difference between the arrival of the start control iteration
timing message and the end control iteration timing message, in the FPGA, over
several thousand iterations of the control loop. The reason for this jitter is
twofold: {\em (a)} one source is the jitter of the execution
time itself (the difference between the minimum and maximum execution time as
shown in Figure \ref{fig::detection}) and {\em (b)} the second source of jitter
is the varying time of message propagation through the PCIe bus. 

Since our testbed was a multicore system, processes running concurrently on
other cores as well as other independent bus masters such as peripherals may
cause interference on the shared interconnect. Although, as we already said, in
a deployed real-time control system, such noise would not be present or at least
bounded. On the other hand, our testbed was essentially an off-the-shelf
installation of Linux running on COTS hardware. Nonetheless, we measured the
typical timing variation caused by the interconnect to be about $0.6$
microseconds, or less than {\em one eighth of the iteration time of a single
control iteration in this case}. Any malicious code that increases the execution
time of the task by more than this amount would be detected by the FPGA timing
monitor. As shown in Table \ref{tab:intr_timing}, we can now detect an intrusion
using the timing-based side channel within $5.7 \mu$s and anything that changes
the timing by $0.6 \mu s$ would be caught. Furthermore, we could also add
multiple timing messages in each control iteration (since the CPU message
overhead is so low) to further reduce the maximum intrusion detection delay. 

\begin{table}[tb]
\begin{center}
\begin{tabular}{|l|r|}
\hline
Measured Quantity & Time ($\mu$s) \\
\hline
\hline
Control Task Execution Time & 4.8 - 5.4  \\
for single iteration	&	\\
\hline
Interconnect Extra Jitter & $\sim$ 0.6 \\
\hline
Enforced Iteration Time & 4.6 - 5.7 \\
\hline
Timing Anomaly Detection Time & 5.7 \\
(for Inverted Pendulum system) & \\
\hline
Vanilla Simplex Anomaly Detection Time &  10,000\\
\hline
Timing Message CPU Overhead & 0.05 \\
\hline
\end{tabular}
\end{center}
\vspace{-\baselineskip}
\caption{Measured Timings during Intrusion Detection}
\label{tab:intr_timing}
\end{table}

In Table \ref{tab:intr_timing}, the control task execution time was obtained from
from the execution time measurements from Section \ref{subsec:timing_results}.
The values in the table are in absolute time which was converted from the cycle
count measurements we performed. Hence, the $4.6 - 5.7 \mu s$ value for the
`Control Task Execution Time' is obtained from the (approx.) $13,000 - 14,000$
cycles that we discussed in Section \ref{subsec:timing_results} and Figure \ref{fig::results_100000}.

Due to the extra jitter caused by the
interconnect, the enforced iteration time is expectedly larger than the measured
control task execution time. The maximum enforced iteration time, $5.7 \mu$s, is
the maximum time the experimental framework can proceed without a timing message
before the safety controller takes over. To state it another way, in the FSM in
Figure \ref{fig::detection}, the runtime value of $MustWait_C$ is $4.6 \mu$s,
and the runtime value of $CanWait_C$ is about $1.1 \mu$s ($mustWait_I$ and
$canWait_I$ are much lower). Given those numbers, the side-channel monitor FSM
will detect a missed timing message within $5.7 \mu$s, {\em i.e.} the detection
time reported. 

We now compare the early detection of malicious code through timing side channels with the behavior of the decision module of the S3A with the situation when it only monitors
the plant state (vanilla Simplex). In the timing side channel version, as discussed above, the maximum time
that can proceed before without valid timing messages is $5.7 \mu$s. For the vanilla Simplex version, we experimentally measured the amount of time needed to detect an intrusion. After taking control of the system, we
immediately tried to destabilize the pendulum by sending a maximum voltage value
in the direction which would most quickly collapse the pendulum (in order to see
a lower bound on the detection time when plant state is monitored alone). In
this experiment, we were able to detect an intrusion after $5$ control
iterations, or $100$ milliseconds. It is clear that the use of timing side channels enables significantly faster detection of security vulnerabilities in real-time control systems: over four orders of magnitude faster than with traditional Simplex.

\section{Limitations}
\label{sec:limitations}

The proposed S3A Architecture is not a silver bullet for intrusion detection in
embedded control systems and does have some practical restrictions which may
limit its applicability. Firstly, in order to use the Secure System Simplex
Architecture in a real system, the system needs to be designed with the
architecture in mind. If there is no way to insert a decision module between the
controller and the plant then the architecture can not be used. While this is a
limitation for some existing systems, we think that the design of future systems
could provision for such techniques; after all it is never a bad idea to
consider security aspects when designing a new system.

One concern regarding the correctness of the approach is making sure that an
attacker cannot easily replicate our side channels. {\em E.g.}, if a processor
instruction count side channel is created by naively sending the current
instruction count value to the the {\tt Monitoring Module} then a malicious
entity could easily store and then replay these values. These types of restrictions
could be overcome with minor modifications to the processor architecture. In
this instance, allowing the FPGA to directly access the instruction
count without involving explicit communication from the CPU would eliminate the
possibility of spoofing.

Additionally, for each side channel, a model of the correct behavior must be
created that would restrict a malicious program. For our timing side channel,
one problem could be that the execution of the task has too much of a difference
between the minimum and maximum execution times to provide real restrictions on
system behavior. While this could be the case in general purpose systems, it is
not very likely in CPS with real-time constraints. Even so, this could be
overcome at runtime by having each timing-behavior-modifying branch point send a
timing message to the FPGA indicating what path was taken. This would permit
an extremely tight bound to be placed on the execution time at the expense of a
more complicated state machine to detect timing anomalies. The construction and
tuning of the timing parameters of the state machine is also currently a
manual process. We believe this could eventually become a more
automatic step in the procedure by performing a compile-time analysis of the
control flow graph of the code -- indicating where to send the timing messages
and using run-time analysis to perform precise timing measurements.

The implementation of the trusted FPGA hardware in our framework must be
correct for the system to be secure. This may seem like we have just moved the
problem over to securing the FPGA system instead of the main system, but this is
not exactly the same for the following reason: the FPGA and Safety Controller
only need to maintain the safety of the plant. The Complex Controller, on the
other hand, can perform useful work with the plant so any upgrades will
be made to the Complex Controller and not to the FPGA's safety logic. The
Complex Controller's timing profile would need to be upgraded but that could
be done in a restricted way to prevent modification to the Safety Controller and
Decision Module. Of course, we should not permit FPGA reconfiguration at runtime
and the trusted hardware platform could even be created on an
Application-Specific Integrated Circuit (ASIC) instead of an FPGA in order to
enforce this.

One issue related to the use of FPGAs in such systems is that sometimes
the complex controller might require the use of complex floating point
calculations and such floating-point computation units are typically not present
on FPGAs since they use up significant area. 
The FPGA in our architecture is used
as a rapid prototype of the trusted simplex component. A deployment 
implementation would likely use a trusted microcontroller along with any 
capabilities (floating point unit) that are needed for the safety controller, 
decision module, and side-channel monitor. Also, as mentioned before, the FPGA
will only host the safety controller that maintains bare functionality. Hence,
it is unlikely that it will need to perform fancy floating point calculations.

Finally, the original Simplex, in general, can only protect the systems from
properties known up front to result in unsafe states. {\em E.g.} in Stuxnet, the malicious
controller would actuate the plant motor for periods at very high frequencies and then for
periods at very low frequencies in order to damage the motors. If the
{\tt Decision Module} was not monitoring this property, such unsafe actuation
would still proceed to the plant.

\section{Related Work}
\label{sec:related}

The closest work to S3A is by Zimmer {\em et. al.}\cite{zimmer10}. They use
worst-case execution time (WCET) information to detect intrusions in hard
real-time systems by instrumenting the tasks and schedulers and periodically checking
whether the execution has gone past the expected WCET values. Our work is more
focused on detecting intrusions in real-time control systems and ensuring that
the plant remains safe even if the intruder is able to bypass all the
detection/security mechanisms. Also, in our work, the system remains safe even
if the intruder gains root privileges to the system -- the work by Zimmer {\em
et. al.} cannot withstand this level of intrusion since an attacker with root
privileges can bypass all the checking mechanisms. Also, our checking/monitoring
is performed by a trusted hardware component that is separate from the main
system thus increasing the overall robustness of the architecture.

The trusted computing engine (TCE) \cite{iyer2008} as well as the
reliability and security engine (RSE) \cite{hardwaresecurity:iyer2007} also use
secure co-processors to execute security-critical code and to monitor the access of
critical data. During setup the security-critical application is loaded on the
TCE and then access to it is monitored during runtime. To detect other security
violations, compile-time analysis is performed to determine the critical data,
the dataflow and what parts of the code are allowed to access this data. At
runtime, RSE monitors all of this information to see if unauthorized instructions/programs
access this critical data. While these techniques could be combined with S3A
(since they are more about intrusion prevention), we don't need to know the
information about what data is critical or even touch the source code. We detect
intrusions by observing the innate characteristics of the program at runtime.

The IBM $4758$ secure co-processor could also be used to perform intrusion
detection \cite{hardwaresecurity:zhang2002}. This work contains a CPU,
separate memory (volatile and non-volatile) along with cryptographic accelerators and comes
wrapped in a tamper-responding secure boundary. The main methods employed for
intrusion detection included checking the system for invariants (one example
was that a normal user's `uid' should never change to root) and detecting
related violations. Also, they used it to execute the virus checking programs
since it couldn't be tampered with. While we could adapt this processor
for use with our architecture, the main difference from S3A lies in the fact
that we employ the inherent characteristics of the program to detect intrusions,
especially in the CPS domain; also coupling with the System Simplex mechanism
increases the robustness of the overall system. 

FlexCore \cite{hardwaresecurity:deng2010} uses a reconfigurable fabric to
implement monitoring and book-keeping functions. It can be used to implement bookkeeping
mechanisms and specific security methods such as array bounds checking,
uninitialized memory checks, dynamic information flow tracking, {\em etc.} in
the reconfigurable hardware. While many of these functionalities could be
implemented in S3A, the main difference with FlexCore lies in the fact that we
{\em (a)} don't need to know what types of attacks are taking place
(as long as it modified the execution time behavior of our code) and {\em (b)}
don't need to analyze the program structure/data as will be the case with
FlexCore.

Pioneer \cite{seshadri2005pioneer} uses sophisticated checksum code and its
execution time information to establish safe remote execution on an untrusted
computer. The checksum code is carefully designed so that any malicious
modification will result in increased execution time that will be detected by
the requesting computer. While their goal is remotely executing arbitrary code
safely on untrusted computers, our goal is to detect the behavioral changes of
known code running on potentially compromised computers.

TVA \cite{garfinkel2003terra} provides guarantees that the software running on
the computer is safe in conjunction with a hardware trusted component (TPM).
While they  also use trusted hardware, our approach differs in that we're not
trying to prevent intrusions or attacks. Our aim is to detect these quickly and
maintain the physical safety of the plant.

Other related work is the use of PRET (precision timed machines) to
detect and protect against side-channel attacks \cite{Liu:EECS-2009-15}. While
their work is focused on preventing attacks based on side-channels we use them
for a benevolent purpose -- to improve the overall security of the system.
\section{Conclusions and Future work}
\label{sec:concl}

In this paper we presented a new framework named Secure System Simplex
Architecture (S3A) that enhances the security and safety of a real-time control
system such as a SCADA plant. We use a combination of trusted hardware,
benevolent side-channels, OS techniques and the intrinsic
real-time nature (and domain-specific characteristics) of such systems to detect
intrusions and prevent the physical plant from being damaged. We were able to detect
intrusions in the system in less than $6\ \mu s$ and changes of less than $0.6\ 
\mu s$ -- time scales that are extremely hard for an intruders to defeat. This
paper also shows that even if an attacker is able to bypass all
security/intrusion detection techniques, the actual plant will remain safe. 
Another important characteristic of these techniques is that there are {\em no
modifications required in the source code}. We believe that the novel techniques
and architecture presented in this paper will significantly increase the
difficulty faced by would-be attackers thus improving the security and overall
safety of such systems.

The intrusion detection capabilities of S3A can be further enhanced by
monitoring multiple side channels and/or improving the predictability of the
system. {\em E.g.}, with the current implementation, the more the system is
predictable, the less will be the jitter measured by the timing analysis,
the tighter can be the execution time range enforced by the Secure Simplex.

For {\bf future work} we plan to investigate other side channels. For instance,
instruction count can be used in S3A so that a deviation in the number of
instructions can be treated as an indication of the existence of malicious code.
Fairly small modifications in the processor could enable trusted hardware
to access the CPU instruction counter, thus enabling an \emph{instruction-based
side channel}.
Finally, a predictable execution model like \emph{PREM} \cite{prem2011}, can also
considerably enhance system predictability and hence, the precision
of timing side channel. In fact, PREM can almost eliminate the jitter
in execution time jitter that results from bus and memory contentions.

\bibliographystyle{ieee}
\bibliography{bak,sibin,ada9x,arch,embedded-software,embedded-system,ieee,mybib,process_variation,power,realtime,regehr,sibin.avionics,sibin.security,sibin.models,yuanxie,zhu,minyoung,heechul.misc,emi}


\end{document}